\documentclass{article}
\usepackage{spconf,amsmath,graphicx}
\usepackage{tabularx,booktabs}

\title{DHASP: DIFFERENTIABLE HEARING AID SPEECH PROCESSING }

\name{Zehai Tu, Ning Ma, Jon Barker}
\address{University of Sheffield, Department of Computer Science, Sheffield, UK\\
\textit{\{ztu3, n.ma, j.p.barker\}@sheffield.ac.uk}}

\begin{document}
\ninept
\maketitle

\begin{abstract}
Hearing aids are expected to improve speech intelligibility for listeners with hearing impairment. An appropriate amplification fitting tuned for the listener's hearing disability is critical for good performance. The developments of most prescriptive fittings are based on data collected in subjective listening experiments, which are usually expensive and time-consuming. 
In this paper, we explore an alternative approach to finding the optimal fitting by introducing a hearing aid speech processing framework, in which the fitting is optimised in an automated way using an intelligibility objective function based on the HASPI physiological auditory model. The framework is fully differentiable, thus can employ the back-propagation algorithm for efficient, data-driven optimisation.
Our initial objective experiments show promising results for noise-free speech amplification, where the automatically optimised processors outperform one of the well recognised hearing aid prescriptions. 
\end{abstract}


{\let\thefootnote\relax\footnote{{© 2021 IEEE. Personal use of this material is permitted. Permission from IEEE must be obtained for all other uses, in any current or future media, including reprinting/republishing this material for advertising or promotional purposes, creating new collective works, for resale or redistribution to servers or lists, or reuse of any copyrighted component of this work in other works.}}}

\vspace*{-6mm}
\section{Introduction}
\label{sec:intro}
\vspace*{-2mm}
It is estimated that approximately 500 million people suffer from hearing impairment around the world and could benefit from hearing aids~\cite{wilson2017global}. Unfortunately, hearing aids work poorly if not  configured correctly for the user, and the fitting process can be expensive and hard to get right.
Inspired by recent advances in deep neural networks for speech processing, this paper proposes a differentiable hearing aid speech processing (DHASP) framework in which a hearing aid processor with trainable parameters can be optimised via back-propagation. Using an existing intelligibility model, the hearing aid is automatically tuned to maximise the predicted intelligibility of the speech signal for a specific individual.

Our aim is to develop an approach that can better individualise a hearing aid for speech perception, while starting with only the standard audiological measurements as a characterisation of the listener. The hope is that this can speed up the fitting process and reduce the number of return visits needed to an audiology clinic. The work presented here is using a relatively simple scenario of speech in quiet conditions and focuses on optimising the hearing aid's frequency-gain amplification. However, the framework can be generalised to more challenging scenarios, with the inclusion of further differentiable hearing aid components, such as noise cancellation and gain control.

In the next section, we provide background for the intelligibility model employed and the baseline fitting. In Section~\ref{sec:method} the proposed hearing aid speech processing framework is described in detail. Section~\ref{sec:experiments} introduces the evaluation framework and the experimental setup. The results are discussed in Section~\ref{sec:results}. Section~\ref{sec:conclusion} concludes the paper and presents some future directions.

\vspace*{-4mm}
\section{Background}
\label{sec:background}
\vspace*{-2mm}

Typically, the hearing-aid's frequency-gain amplification is set based on the listener's measured pure-tone hearing ability using a standardised mapping. Early hearing aid fitting prescriptions, including the National Acoustic Laboratories Revised (NAL-R) formula~\cite{byrne1986national}, aim to maximise speech intelligibility for a specified loudness level, thus the desired output is an optimal frequency response formula. With the introduction of commonly used wide dynamic range compression, recent prescriptions, including NAL-NL1, NAL-NL2~\cite{byrne2001nal, keidser2011nal} and CAMEQ, CAMEQ-2HF~\cite{moore1999use, moore2010development}, give gain-frequency responses to take loudness into consideration. In this paper, we focus on speech at a fixed level, and so for a baseline, we use the NAL-R prescription, which is widely recognised, suitable for  a wide range of hearing disabilities and open-sourced.

The key to our approach is to leverage highly-developed speech intelligibility models that can predict the performance of impaired and corrected hearing.
Such models have been critical for the development of the hearing aid algorithms and the improvement of the hearing aid prescriptions. The speech intelligibility index (SII)~\cite{ansi1997s3}
has been used as the basis for the development of most widely used hearing aid prescriptions. SII predicts intelligibility by sum up the products of the audibility of each frequency band and the corresponding importance factors, and takes hearing loss into consideration by incorporating a distortion factor and audibility reduction.  More sophisticated intelligibility models which take hearing loss into account have been proposed recently.  In this paper we employ the hearing-aid speech perception index (HASPI) developed by Kates and Arehart~\cite{kates2014hearing} which uses a physiological auditory model which incorporates the characteristics of hearing impairment~\cite{kates2013auditory}.

Related to our work, Sch\"{a}dler et al.~\cite{schadler2018objective} proposed to use an automatic speech recogniser to predict the speech-in-noise recognition performance of hearing impaired listeners. The speech to reverberation modulation energy ratio was also tailored to hearing impaired instruments to predict both speech quality and intelligibility~\cite{falk2013non}. 
However, to the best of our knowledge, little work has been done to automatically optimise hearing aid processing against an intelligibility model.

\vspace*{-4mm}
\section{Method}
\label{sec:method}
\vspace*{-2mm}

\begin{figure}[t!]
  \centering
  \includegraphics[width=\linewidth]{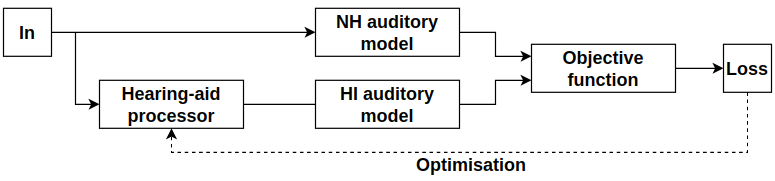}
  \caption{System diagram of the proposed DHASP framework.}
  \label{fig:DHASP}
\end{figure}

Fig.~\ref{fig:DHASP} shows the system diagram of the proposed DHASP framework, which consists of a hearing-aid processor, a normal hearing (NH) auditory model, a hearing impaired (HI) auditory model, and an objective function. A clean signal is processed by the NH model to obtain the reference output. Meanwhile, the same signal is firstly enhanced by the hearing aid processor with trainable parameters, and then processed by the hearing impaired auditory model to obtain the processed output. The difference between the reference and the processed outputs is measured by the objective function. All components in the framework are differentiable so that the hearing-aid processor can be optimised via back-propagation. The components within the DHASP are described in detail below.

\subsection{Differentiable hearing aid processors}
To follow the same setting as NAL-R, differentiable finite impulse response (FIR) filters are used to implement frequency dependent amplification.
FIR filters are used for hearing-aid processing by convolving input signals. These hearing-aid processing filters are determined by a limited number of trainable parameters which represent the gains at certain frequencies. These frequencies are the same as those used in the listeners' audiograms in this work, but can be customised in any scale. Linear interpolation is applied to the parameters to obtain the frequency responses of the filters. Inverse Fourier transform is then used followed by a Hann window to achieve the FIR filters in the time domain.

\subsection{Differentiable auditory model}

\begin{figure}[t!]
  \centering
  \includegraphics[width=\linewidth]{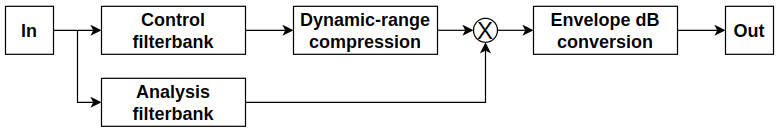}
  \caption{Structure of the differentiable auditory model.}
  \label{fig:auditory_model}
\end{figure}

The workflow of the differentiable auditory model shown in Fig.~\ref{fig:auditory_model} is mainly adapted from the auditory processing used in HASPI~\cite{kates2013auditory}. The model operates at 24\,kHz and depends on the auditory thresholds given by the listener's audiogram at [250, 500, 1000, 2000, 4000, 6000] Hz. The auditory thresholds are set to zeros for normal hearing model. Two groups of filterbanks, the dynamic-range compression, and a dB conversion process, are used to simulate the mechanisms in human audition considering the impact of hearing impairment. In contrast to the auditory model in~\cite{kates2013auditory}, which uses infinite impulse response filters (IIR), the proposed DHASP framework employs FIR filters to avoid expensive recursive computation. The middle ear component and the inner-hair cell adaptation process in the HASPI model are not included for the same reason. The influence of the signal intensity on the analysis filter bank included in the HASPI model is not considered because of the difficulty in the differentiation implementation. All parameter settings used in the differentiable auditory model are the same as the model used in HASPI.

\subsubsection{Analysis filterbank}
The analysis filter bank consists of a total of $I = 32$ fourth-order FIR gammatone filters~\cite{cooke2005modelling}. The $i^{th}$ filter $h_{a}^{(i)}$ of the analysis filterbank is expressed as: 
\begin{equation}
    h_{a}^{(i)}(t) = A_{a}^{(i)} t^{\left(N^{(i)}-1\right)} e^{-2 \pi b_{a}^{(i)} t}\cos \left(2 \pi f_{a}^{(i)} t\right),
\end{equation}
where $A_{a}^{(i)}$ is the amplitude required to normalise the frequency response of the filter; 
$b_{a}^{(i)}$ and $f_{a}^{(i)}$ are the bandwidth and the centre frequency of the filter, respectively~\cite{loweimi2019learning}; $N{(i)}$ is the order of the filter which is set as $4$ in the model. The centre frequencies $f_{a}$ are in the Mel scale covering the range from 80\,Hz to 8\,kHz. The bandwidths $b_{a}^{NH}$ are in the equivalent rectangular bandwidth (ERB) scale~\cite{moore1983suggested} for the normal hearing model. To approximate the behaviour that the auditory filter bandwidths increase along with the hearing loss~\cite{moore1999inter}, the bandwidths $b_{a}^{HL}$ of the hearing loss model is expressed as:
\begin{equation}
    b_{a}^{HL} = \left(1 + attn_{o}/50 + 2(attn_{o}/50)^{6}\right)b_{a}^{NH},
\end{equation}
where $attn_{o}$ is the hearing loss for outer-hair cells in dB, with a maximum attenuation of 50 dB \cite{kates2013auditory}.

\subsubsection{Control filterbank}

Another group of fourth-order FIR gammatone filters are used as control filterbank to simulate the two-tone suppression mechanism in the cochlea~\cite{heinz2001auditory, bruce2003auditory}. The bandwidths of the control filters correspond to the maximum bandwidth allowed in the model, i.e. 50\,dB attenuation for outer-hair cell. The control filters are set wider so that they could reduce the gain of a signal outside the bandwidth of the analysis filters but still within the control filters~\cite{kates2013auditory}. Each center frequency of the control filter $f_{c}^{(i)}$ is shifted higher relative to the center frequency of the corresponding analysis filter $f_{a}^{(i)}$ using a human frequency-position function~\cite{greenwood1990cochlear}:
\begin{equation}
    f_{c}^{(i)} = 165.4(10 ^ {(1 + s)\log_{10}\left(1 + f_{c}^{(i)} / 165.4 \right)} - 1),
\end{equation}
where $s$ is the shift fraction which is set as 0.02 in this model.

\subsubsection{Dynamic-range compression}

The dynamic-range compression is simulated following the control filtering. The input to the compression rule is each control signal envelope $E_{c}^{(i)}(n)$ in dB. The compression gain $G^{(i)}(n)$ in dB is computed as:
\begin{equation}
    G^{(i)}(n) = - attn_{o} - (1 - 1 / CR) (\theta_{low} - \hat{E}_{c}^{(i)}(n)),
\end{equation}
where:
\begin{equation}
    \hat{E}_{c}^{(i)}(n) = max(\theta_{low}, (min(E_{c}^{(i)}(n), \theta_{high})).
\end{equation}
$\theta_{low}$ is the lower threshold set as $(attn_{o} + 30)$ dB; $\theta_{high}$ is set as 100\,dB in the model; and $CR$ is the compression ratio which is 1.25 at 80\,Hz and linearly increases to 3.5 at 8\,kHz for the normal hearing model. This compression behaviour is consistent with the psychophysical estimates of dynamic-range compression in the human auditory system~\cite{moore1999inter}. Increasing outer-hair cell damage leads to the reduction of compression ratio. The maximum damage gain $G_{max_{o}}$ is set as 14 dB for the compression ratio of 1.25 at 80\,Hz, and as 50\,dB for the compression ratio at 8\,kHz. The outer-hair cell threshold is set as $1.25 G_{max_{o}}$. If the hearing loss indicated by the audiogram is greater than the outer-hair threshold, $attn_{o}$ is set as $G_{max_{o}}$ and inner-hair cell loss $attn_{i}$ is set as the remaining loss. On the contrary, $attn_{o}$ and $attn_{i}$ are set as 80\% and 20\% of the total loss, respectively. The compression gain $G^{(i)}(n)$ is then converted into the linear scale, and applied to the corresponding output of the analysis filtering.

\begin{figure*}[t!]
    \centering
    \includegraphics[width=0.95\textwidth]{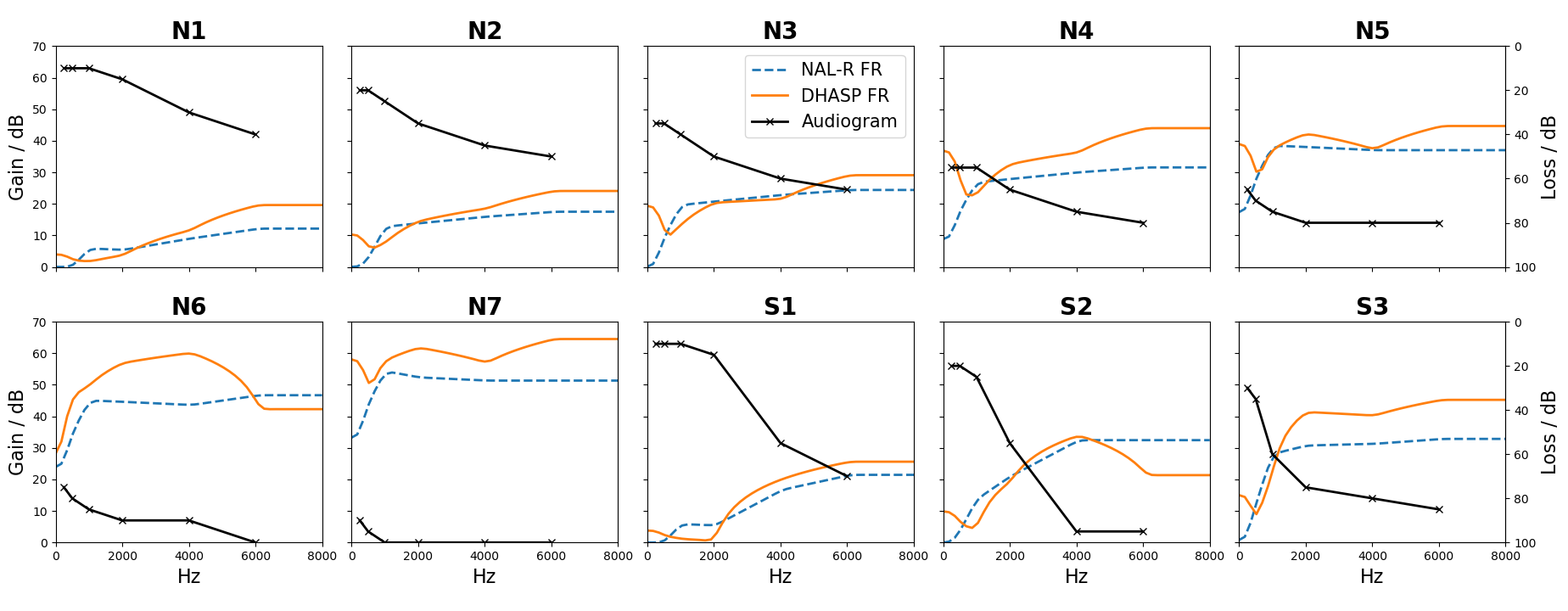}
            \vspace*{-5mm}
    \caption{Standard audiograms are presented as the solid curves with cross markers. The hearing losses are capped at 100\,dB. The dashed and solid curves represented the frequency responses of the NAL-R prescription filters and the trained DHASP filters, respectively.}
    \label{fig:results_audiogram}
\end{figure*}

\subsubsection{Envelope dB conversion}
The compressed analysis envelope is converted into dB at this stage. The inner-cell hair loss attenuation $attn_{i}$ is then added to the converted envelope.

\subsection{Objective function}
The reference envelope $E^{(i)}_{r}(n)$ and processed envelope $E_{p}^{(i)}(n)$ processed by the normal hearing and the hearing loss model, respectively, are smoothed using a 16\,ms Hann window with 50\% overlapping. Given the smoothed envelopes $E^{(i)}_{r}(m)$ and $E_{p}^{(i)}(m)$, the objective function consists of a cepstral correlation measure function~\cite{kates2014hearing} and an energy control function. A set of half-cosine basis functions $b_{j}(i)$ are used to compute the cepstral sequences:
\begin{equation}
    C^{(j)}_{r}(m) = \sum_{i=1}^{I}b_{j}(i)E^{(i)}_{r}(m),
\end{equation}
\begin{equation}
    C^{(j)}_{p}(m) = \sum_{i=1}^{I}b_{j}(i)E^{(i)}_{p}(m),
\end{equation}
where:
\begin{equation}
    b_{j}(i) = cos[(j - 1)\pi i / (I - 1)].
\end{equation}
These basis functions are similar to the principal components for the short-time spectra of speech~\cite{zahorian1981principal} and have been used for consonant and vowel recognition~\cite{nossair1991dynamic, zahorian1993spectral}. The normalised correlation is then expressed as:
\begin{equation}
    R(j) = \frac{\sum_{m=0} C^{(j)}_{r}(m) C^{(j)}_{p}(m)}{\sqrt{\sum_{m=0} (C^{(j)}_{r}(m))^2}\sqrt{\sum_{m=0} (C^{(j)}_{p}(m))^2}}.
\end{equation}
The final cepstral correlation is the average of $R(2)$ to $R(6)$.

To prevent the over-amplification of the trained hearing-aid processors, which brings discomfort to listeners, an energy control loss is introduced to constrain the processed envelope energy if it is higher than the corresponding reference envelope energy:
\begin{equation}
    L_{e}^{(i)} = \sum_{m \in S }(E^{(i)}_{p}(m) - E^{(i)}_{r}(m)), 
\end{equation}
where:
\begin{equation}
    S = \{ m \mid E^{(i)}_{p}(m) - E^{(i)}_{r}(m) > 0\}.
\end{equation}
Overall, the objective function used is expressed as:
\begin{equation}
    L = - \frac{1}{5} \sum_{j=2}^{6}R(j) + \alpha \sum_{i}L_{e}^{(i)},
\end{equation}
where $\alpha$ is the energy loss weight, which is tuned empirically.

\begin{figure*}[h!]
    \centering
    \includegraphics[width=\textwidth]{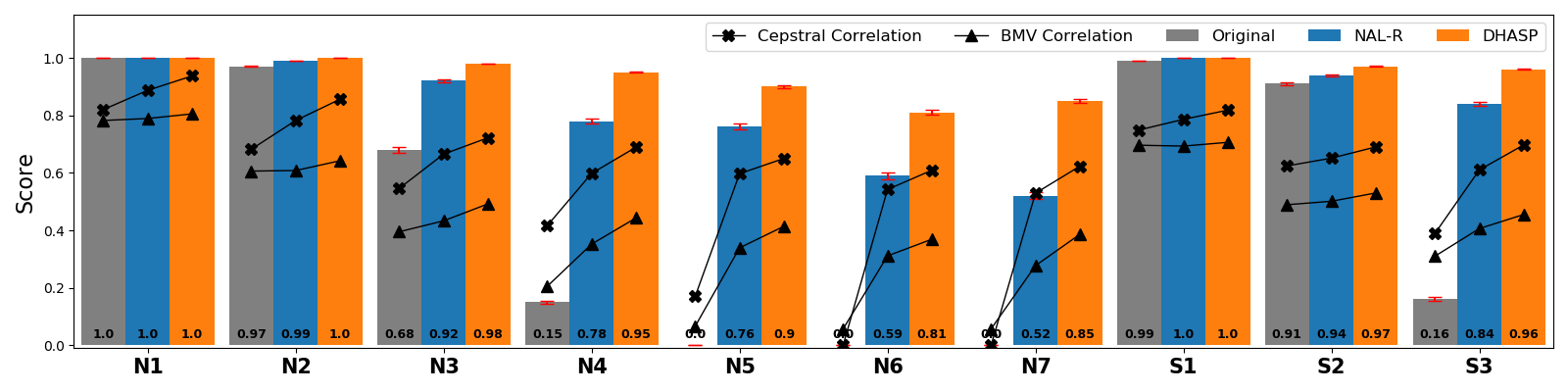}
            \vspace*{-5mm}
    \caption{HASPI intelligibility scores of the original, NAL-R processed, and DHASP processed signals. The error bars indicate the standard error of the mean utterance intelligibility scores in the test dataset. The curves with the cross and triangle marks show the corresponding cepstral correlation scores and the BMV correlation scores.}
    \label{fig:results_barchart}
\end{figure*}

\section{Experiments}
\label{sec:experiments}
\subsection{Evaluation}
HASPI is used to evaluate the performance of the proposed framework. HASPI is based on the auditory model proposed in~\cite{kates2013auditory}, and it is designed to predict the speech intelligibility for hearing impaired listeners. Both cepstral correlation $C_{C}$ and basilar membrane vibration (BMV) correlation $C_{B}$, based on the envelopes and the temporal fine structures, respectively, are taken into consideration. The HASPI intelligibility score $H$ is computed as a linear combination followed by a nonlinear scaling function:
\begin{equation}
    H = \frac{1}{1 + e^{-(14.817C_{C} + 4.616C_{B} - 9.047)}}
\end{equation}
HASPI has been validated in multiple databases comprising intelligibility scores~\cite{kates2005coherence, souza2013exploring, arehart2013relationship}.

NAL-R prescription is used as the baseline system. It prescribes a gain frequency curve given an audiogram. The hearing losses at [250, 500, 1000, 2000, 4000, 6000] Hz are used for the frequency response derivation to be consistent with the proposed framework. A FIR filter is then designed as the hearing aid processor given the frequency response curve.

\subsection{Audiogram database}
10 standard audiograms, which cover a range of common audiograms  in clinical practice~\cite{bisgaard2010standard}, are used to evaluate the proposed framework and are shown in Fig.~\ref{fig:results_audiogram} as the solid curves with crossing marks. N1 to N7 represent hearing impaired listeners with the flat and moderately sloping audiograms, and S1 to S3 represent the steep sloping group. The audiograms are ranked according to the hearing loss severity. As HASPI has a maximum hearing loss limit, the audiograms used in this work are capped at deficits of 100\,dB.

\subsection{Experimental setup}
DHASP is implemented using a popular automatic differentiation tool PyTorch, and is trained and evaluated on the TIMIT dataset~\cite{garofolo1993darpa}. The training set consists of utterances from 462 speakers while utterances from 50 speakers are used as the validation set. Utterances of the remaining 24 speakers are used as the final evaluation test set. In both training and evaluation, the input signal is normalised so that its root-mean-square equals one and is regarded as 65\,dB SPL to mimic everyday conversational speech. Utterance segments of 0.5-second long are randomly sampled as the input signals during training. The processors are trained with a batch size of 128 for 4000 epochs using the Adam optimiser~\cite{kingma2014adam} and a learning rate of 0.001. Six trainable parameters which represent the frequency response gains of the processors at [250, 500, 1000, 2000, 4000, 6000] Hz are optimised. The parameters are all initialised to 1\,dB for audiograms N1 to N6 and S1 to S3. For profound loss such as audiogram N7, the low gain initialisation leads to vanishing gradients. Therefore the parameters are initialised to 50\,dB in the experiment. The energy loss coefficient $\alpha$ is set to 5e-5.

\section{Results and discussions}
\label{sec:results}

The frequency responses of the optimised filters by DHASP and the NAL-R prescription filters are shown in Fig.~\ref{fig:results_audiogram} as the solid and dashed curves, respectively. 
Across all the standard audiograms the amplification provided by the optimised filters broadly follows the frequency response patterns of the NAL-R filters.
In general, the optimised filters amplify the input signals more in the frequency region below 500\,Hz. For audiograms with mild and moderate high frequency loss (N1-N5), the proposed filters have higher gains in the high frequency area. On the contrary, the proposed filters amplify less in the high frequency when the loss is severe as shown in N6 and S2. As the information in the high frequency is almost not recoverable due to the profound loss, the amplification in that area would not make a significant difference. DHASP ensures stable convergence for the training of all audiograms, while the convergence time increases along with the severity of the hearing loss.

Fig.~\ref{fig:results_barchart} shows the HASPI scores, including intelligibility scores $H$, cepstral correlation $C_{C}$ and simulated BMV correlation $C_{B}$, of the unprocessed original signals, NAL-R processed signals, and DHASP-processed signals. Both processed signals had higher intelligibility scores than the unprocessed signals. The filters optimised by the DHASP framework achieved higher HASPI scores than the NAL-R prescription filters, with improvements significant across all the audiogram conditions [paired $t$-test, $p<.005$]. 
With increased hearing loss severity, the advantages of the proposed optimised processors are more significant compared to the NAL-R prescription. The variation of the intelligibility scores across all the utterances in the test dataset indicates that DHASP can achieve better performance with good consistency. It is not surprising that the cepstral scores of the optimised filters are higher than the NAL-R ones because the objective focuses on the envelope correlation. However, the optimised filters also consistently achieve higher BMV correlation scores.

\section{Conclusions}
\label{sec:conclusion}
This paper has proposed the DHASP framework, which is fully differentiable, therefore can optimise the hearing aid processors with the back-propagation algorithm. According to the HASPI metric, the processors optimised by the DHASP framework outperform the NAL-R prescription processors given a range of standard audiograms. The objective function used in the DHASP is a combination of an auditory based intelligibility model and an energy constraint function. It can be improved with further findings on the auditory periphery mechanisms, and refined with more detailed constraints including hearing aid hardware implementation limitations.
With the introduction of machine learning techniques, DHASP has the potential to help the further fine-tuning of the hearing aid fitting as well. Moreover, this framework can also be used for the optimisations of more powerful models like deep neural networks due to the differentiable characteristic. Thus, it has the potential to help tackle various complex challenges, such as speech denoising and separation, for hearing impaired listeners.

So far the DHASP framework has been evaluated with a purely objective metric. Subjective evaluation with hearing impaired listeners is planned for the future. Comparisons with prescriptions offering frequency-gain responses, such as NAL-NL2 and CAMEQ-2HF, will also be conducted.
As the method proposed is data-driven, the framework puts strong prior assumption on the data used for training. Additional speech datasets with more variation than TIMIT (e.g., conversational speech rather than read speech) can be evaluated in further investigation.

\pagebreak
\bibliographystyle{IEEEbib}
\bibliography{Paper}

\begin{thebibliography}{10}

\bibitem{wilson2017global}
Blake~S Wilson, Debara~L Tucci, Michael~H Merson, and Gerard~M O'Donoghue,
\newblock ``Global hearing health care: new findings and perspectives,''
\newblock {\em The Lancet}, vol. 390, no. 10111, pp. 2503--2515, 2017.

\bibitem{byrne1986national}
Denis Byrne and Harvey Dillon,
\newblock ``The national acoustic laboratories'({NAL}) new procedure for
  selecting the gain and frequency response of a hearing aid,''
\newblock {\em Ear and hearing}, vol. 7, no. 4, pp. 257--265, 1986.

\bibitem{byrne2001nal}
Denis Byrne, Harvey Dillon, Teresa Ching, Richard Katsch, and Gitte Keidser,
\newblock ``{NAL-NL1} procedure for fitting nonlinear hearing aids:
  characteristics and comparisons with other procedures.,''
\newblock {\em Journal of the American academy of audiology}, vol. 12, no. 1,
  2001.

\bibitem{keidser2011nal}
Gitte Keidser, Harvey Dillon, Matthew Flax, Teresa Ching, and Scott Brewer,
\newblock ``The {NAL-NL2} prescription procedure,''
\newblock {\em Audiology research}, vol. 1, no. 1, 2011.

\bibitem{moore1999use}
BCJ Moore, BR~Glasberg, and MA~Stone,
\newblock ``Use of a loudness model for hearing aid fitting: {III}. a general
  method for deriving initial fittings for hearing aids with multi-channel
  compression,''
\newblock {\em British Journal of Audiology}, vol. 33, no. 4, pp. 241--258,
  1999.

\bibitem{moore2010development}
Brian~CJ Moore, Brian~R Glasberg, and Michael~A Stone,
\newblock ``Development of a new method for deriving initial fittings for
  hearing aids with multi-channel compression: {CAMEQ2-HF},''
\newblock {\em International Journal of Audiology}, vol. 49, no. 3, pp.
  216--227, 2010.

\bibitem{ansi1997s3}
ANSI ANSI,
\newblock ``S3. 5-1997, methods for the calculation of the speech
  intelligibility index,''
\newblock {\em New York: American National Standards Institute}, vol. 19, pp.
  90--119, 1997.

\bibitem{kates2014hearing}
James~M Kates and Kathryn~H Arehart,
\newblock ``The hearing-aid speech perception index ({HASPI}),''
\newblock {\em Speech Communication}, vol. 65, pp. 75--93, 2014.

\bibitem{kates2013auditory}
James Kates,
\newblock ``An auditory model for intelligibility and quality predictions,''
\newblock in {\em Proceedings of Meetings on Acoustics ICA2013}. Acoustical
  Society of America, 2013, vol.~19, p. 050184.

\bibitem{schadler2018objective}
Marc~R Sch{\"a}dler, Anna Warzybok, and Birger Kollmeier,
\newblock ``Objective prediction of hearing aid benefit across listener groups
  using machine learning: Speech recognition performance with binaural
  noise-reduction algorithms,''
\newblock {\em Trends in hearing}, vol. 22, pp. 2331216518768954, 2018.

\bibitem{falk2013non}
Tiago~H Falk, Stefano Cosentino, Joao Santos, David Suelzle, and Vijay Parsa,
\newblock ``Non-intrusive objective speech quality and intelligibility
  prediction for hearing instruments in complex listening environments,''
\newblock in {\em 2013 IEEE International Conference on Acoustics, Speech and
  Signal Processing}. IEEE, 2013, pp. 7820--7824.

\bibitem{cooke2005modelling}
Martin Cooke,
\newblock {\em Modelling auditory processing and organisation}, vol.~7,
\newblock Cambridge University Press, 2005.

\bibitem{loweimi2019learning}
Erfan Loweimi, Peter Bell, and Steve Renals,
\newblock ``On learning interpretable {CNN}s with parametric modulated
  kernel-based filters.,''
\newblock in {\em INTERSPEECH}, 2019, pp. 3480--3484.

\bibitem{moore1983suggested}
Brian~CJ Moore and Brian~R Glasberg,
\newblock ``Suggested formulae for calculating auditory-filter bandwidths and
  excitation patterns,''
\newblock {\em The journal of the acoustical society of America}, vol. 74, no.
  3, pp. 750--753, 1983.

\bibitem{moore1999inter}
Brian~CJ Moore, Deborah~A Vickers, Christopher~J Plack, and Andrew~J Oxenham,
\newblock ``Inter-relationship between different psychoacoustic measures
  assumed to be related to the cochlear active mechanism,''
\newblock {\em The Journal of the Acoustical Society of America}, vol. 106, no.
  5, pp. 2761--2778, 1999.

\bibitem{heinz2001auditory}
Michael~G Heinz, Xuedong Zhang, Ian~C Bruce, and Laurel~H Carney,
\newblock ``Auditory nerve model for predicting performance limits of normal
  and impaired listeners,''
\newblock {\em Acoustics Research Letters Online}, vol. 2, no. 3, pp. 91--96,
  2001.

\bibitem{bruce2003auditory}
Ian~C Bruce, Murray~B Sachs, and Eric~D Young,
\newblock ``An auditory-periphery model of the effects of acoustic trauma on
  auditory nerve responses,''
\newblock {\em The Journal of the Acoustical Society of America}, vol. 113, no.
  1, pp. 369--388, 2003.

\bibitem{greenwood1990cochlear}
Donald~D Greenwood,
\newblock ``A cochlear frequency-position function for several species—--29
  years later,''
\newblock {\em The Journal of the Acoustical Society of America}, vol. 87, no.
  6, pp. 2592--2605, 1990.

\bibitem{zahorian1981principal}
Stephen~A Zahorian and Martin Rothenberg,
\newblock ``Principal-components analysis for low-redundancy encoding of speech
  spectra,''
\newblock {\em The Journal of the Acoustical society of America}, vol. 69, no.
  3, pp. 832--845, 1981.

\bibitem{nossair1991dynamic}
Zaki~B Nossair and Stephen~A Zahorian,
\newblock ``Dynamic spectral shape features as acoustic correlates for initial
  stop consonants,''
\newblock {\em The Journal of the Acoustical Society of America}, vol. 89, no.
  6, pp. 2978--2991, 1991.

\bibitem{zahorian1993spectral}
Stephen~A Zahorian and Amir~Jalali Jagharghi,
\newblock ``Spectral-shape features versus formants as acoustic correlates for
  vowels,''
\newblock {\em The Journal of the Acoustical Society of America}, vol. 94, no.
  4, pp. 1966--1982, 1993.

\bibitem{kates2005coherence}
James~M Kates and Kathryn~H Arehart,
\newblock ``Coherence and the speech intelligibility index,''
\newblock {\em The journal of the acoustical society of America}, vol. 117, no.
  4, pp. 2224--2237, 2005.

\bibitem{souza2013exploring}
Pamela~E Souza, Kathryn~H Arehart, James~M Kates, Naomi~BH Croghan, and Namita
  Gehani,
\newblock ``Exploring the limits of frequency lowering,''
\newblock {\em Journal of Speech, Language, and Hearing Research}, 2013.

\bibitem{arehart2013relationship}
Kathryn Arehart, Pamela Souza, Thomas Lunner, Michael Syskind~Pedersen, and
  James Kates,
\newblock ``Relationship between distortion and working memory for digital
  noise-reduction processing in hearing aids,''
\newblock in {\em Proceedings of Meetings on Acoustics ICA2013}. Acoustical
  Society of America, 2013, vol.~19, p. 050084.

\bibitem{bisgaard2010standard}
Nikolai Bisgaard, Marcel~SMG Vlaming, and Martin Dahlquist,
\newblock ``Standard audiograms for the {IEC} 60118-15 measurement procedure,''
\newblock {\em Trends in amplification}, vol. 14, no. 2, pp. 113--120, 2010.

\bibitem{garofolo1993darpa}
John~S Garofolo, Lori~F Lamel, William~M Fisher, Jonathan~G Fiscus, and David~S
  Pallett,
\newblock ``{DARPA TIMIT} acoustic-phonetic continuous speech corpus {CD-ROM}.
  {NIST} speech disc 1-1.1,''
\newblock {\em STIN}, vol. 93, pp. 27403, 1993.

\bibitem{kingma2014adam}
Diederik~P Kingma and Jimmy Ba,
\newblock ``Adam: A method for stochastic optimization,''
\newblock {\em ar{X}iv preprint ar{X}iv:1412.6980}, 2014.

\end{thebibliography}

\end{document}